\documentclass[aps,pra,twocolumn,showpacs,superscriptaddress,shortbibliography]{revtex4-2}
\usepackage{graphicx} 
\usepackage{amsmath}
\usepackage{graphicx,epstopdf}
\usepackage{gensymb}
\usepackage[dvipsnames]{xcolor}
\usepackage{footnote}
\usepackage{multibib}

\newcommand{\be}{\begin{equation}}
	\newcommand{\ee}{\end{equation}}
\newcommand{\bea}{\begin{eqnarray}}
	\newcommand{\eea}{\end{eqnarray}}
\newcommand{\bse}{\begin{subequations}}
	\newcommand{\ese}{\end{subequations}}

\graphicspath{{../}}
\usepackage{color}
\usepackage[colorlinks,bookmarks=false,citecolor=darkblue,linkcolor=red,urlcolor=blue]{hyperref}

\definecolor{darkred}{rgb}{0.7,0.0,0.0}

\definecolor{darkblue}{rgb}{0,0.02,0.45}

\definecolor{darkgreen}{rgb}{0.02,0.45,0.0}

\definecolor{violet}{rgb}{0.8,0.2,0.6}

\begin{document}
\title{Ground-state properties of the double trillium lattice antiferromagnet KBaCr$_2$(PO$_4$)$_3$}
\author{R. Kolay}
\affiliation{School of Physics, Indian Institute of Science Education and Research Thiruvananthapuram-695551, India}
\author{Qing-Ping Ding}
\author{Y. Furukawa}
\affiliation{Ames National Laboratory and Department of Physics and Astronomy, Iowa State University, Ames, Iowa 50011, USA}
\author{A. A. Tsirlin}
\affiliation{Felix Bloch Institute for Solid-State Physics, Leipzig University, 04103 Leipzig, Germany}
\author{R. Nath}
\email{rnath@iisertvm.ac.in}
\affiliation{School of Physics, Indian Institute of Science Education and Research Thiruvananthapuram-695551, India}
\date{\today}
	
\begin{abstract}
Trillium lattices formed by corner-shared triangular units are the platform for magnetic frustration in three dimensions. Herein, we report structural and magnetic properties of the Cr-based double trillium lattice material KBaCr$_2$(PO$_4$)$_3$ studied by x-ray diffraction, magnetization, heat capacity, thermal conductivity, and $^{31}$P nuclear magnetic resonance (NMR) measurements complemented by density-functional band-structure calculations. Heat capacity and $^{31}$P NMR measurements reveal the magnetic transition at $T_{\rm N1} \simeq 13.5$~K in zero field followed by another transition at $T_{\rm N2} \simeq 7$~K in weak applied fields. The NMR sublattice magnetization confirms that the transition at $T_{\rm N1}$ is 3D in nature. The $^{31}$P spin-lattice relaxation rate in the ordered state follows the $T^3$ behavior indicative of the two-magnon Raman process. The spin lattice of KBaCr$_2$(PO$_4$)$_3$ comprises two crystallographically nonequivalent ferromagnetic sublattices that are coupled antiferromagnetically, thus eliminating frustration in this trillium network.
\end{abstract}

\maketitle

\section{Introduction}
Frustrated magnets where configurations of localized magnetic moments do not satisfy pair-wise interactions simultaneously often have large ground-state degeneracy and realize exotic phases at low temperatures. The source of frustration could be either spin-lattice geometry with triangular loops or competing interactions between nearest neighbors (NN) and further neighbors~\cite{Ramirez453}. Owing to the ground-state degeneracy, these quantum magnets evade the conventional magnetic long-range order (LRO) and serve as prime candidates to foster various nontrivial ground states such as quantum spin liquid (QSL), spin ice, and unconventional spin glass~\cite{Zhou025003,Binder801,Bramwell2001}. Over the years, significant attention has been focused in realizing frustrated magnets in two dimensions within the triangular~\cite{li2020b} and kagome~\cite{Helton107204} geometries, whereas pyrochlore and hyperkagome lattices have been the primary frustration motifs in three dimensions (3D)~\cite{Gao1052,Plumb54,Chillal2348,Shockley047201}. 

The family of frustrated 3D magnets was recently augmented by the trillium lattice formed by corner-shared triangular motifs with six nearest neighbors. A distinctive feature of this geometry is its chiral nature that has interesting implications in itinerant systems, such as MnSi, FeGe, and MnGe with skyrmion magnetic phases~\cite{muehlbauer2009,Nakajima2017,Yu106} and topological Hall effect~\cite{neubauer2009,Kanazawa156603}. Na[Mn(HCOO)$_3$], the insulating metal-organic-framework compound with the trillium geometry, exhibits a pseudo-plateau at $1/3$ of the saturation magnetization and reveals an unusual 2-\textbf{k} magnetic ground state~\cite{Bulled177201}. A few oxide-based trillium-lattice compounds belonging to the langbeinite structure type have been investigated too. K$_2$Ni$_2$(SO$_4$)$_3$ with $S=1$ moments of Ni$^{2+}$ was shown to feature the field-induced QSL state~\cite{Ivica157204} and a spinon continuum witnessed by inelastic neutron scattering~\cite{Yao146701}. Another langbeinite-type compound KSrFe$_2$(PO$_4$)$_3$ was shown to evade long-range order already in zero field~\cite{Boya101103}, whereas K$_2$CrTi(PO$_4)_3$ with the mixture of magnetic Cr$^{3+}$ and nonmagnetic Ti$^{4+}$ develops two magnetic transitions and non-trivial spin dynamics~\cite{khatua2024}. 

\begin{figure}
\includegraphics[width=\columnwidth] {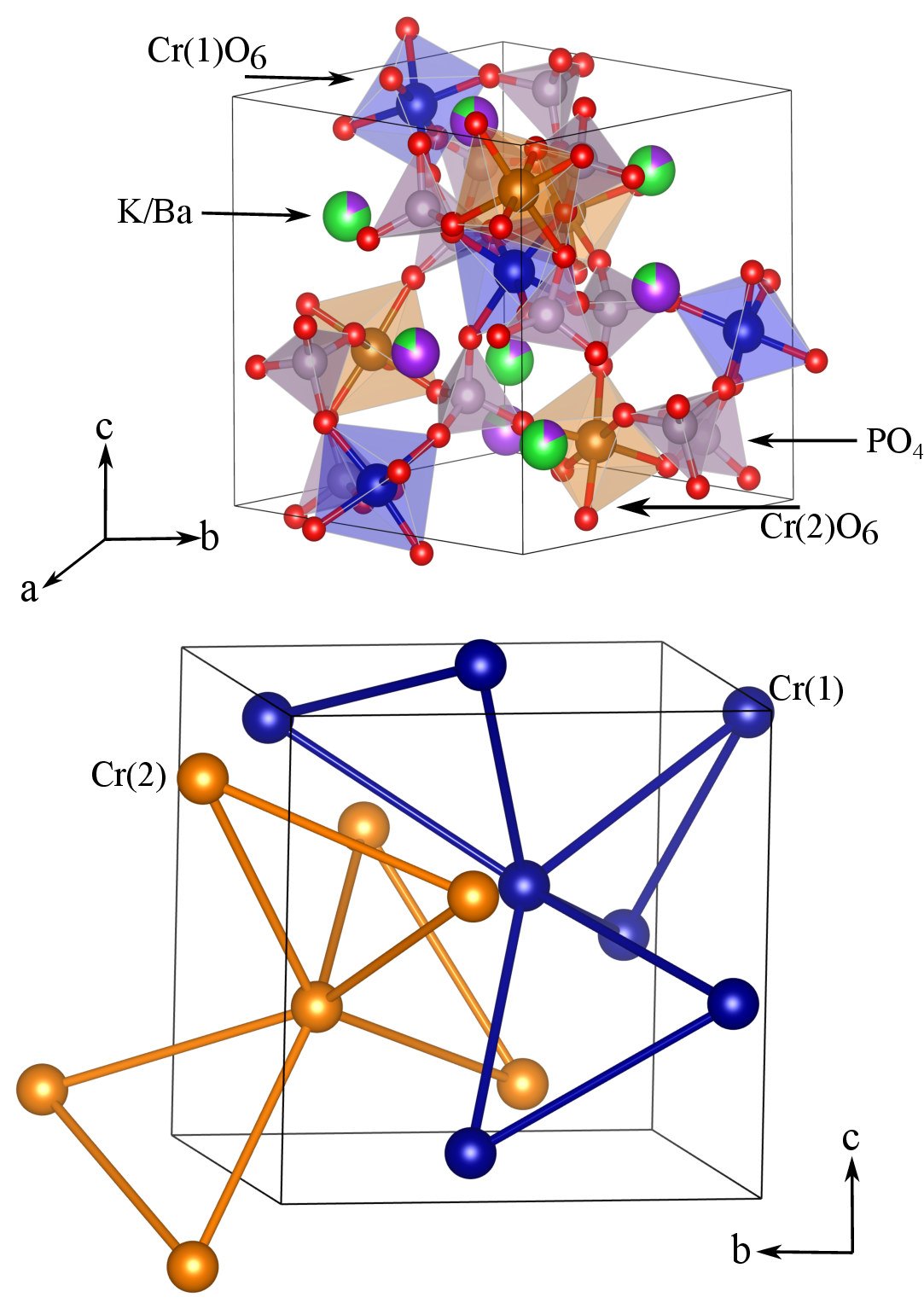}
\caption{\label{Fig1}(a) Crystal structure of KBaCr$_2$(PO$_4$)$_3$ featuring the Cr(1)O$_6$ and Cr(2)O$_6$ octahedra along with the PO$_4$ tetrahedral units. (b) Two interpenetrating trillium lattices formed by the Cr(1) and Cr(2) sites.}
 \end{figure}
In this work, we report magnetism of the KBaCr$_2$(PO$_4$)$_3$ compound~\cite{Battle21} that lacks the effect of magnetic dilution present in K$_2$CrTi(PO$_4)_3$. Interestingly, our data reveal several similarities to the diluted case, including two magnetic transitions despite the absence of frustration in this two-sublattice antiferromagnet. KBaCr$_2$(PO$_4$)$_3$ belongs to the langbeinite structure type (space group $P2_13$). Two Cr$^{3+}$ sites form interpenetrating trillium lattices, resulting in the double-trillium-lattice structure [see Fig.~\ref{Fig1}].

\section{Methods}
\begin{figure}
	\includegraphics[width=\columnwidth]{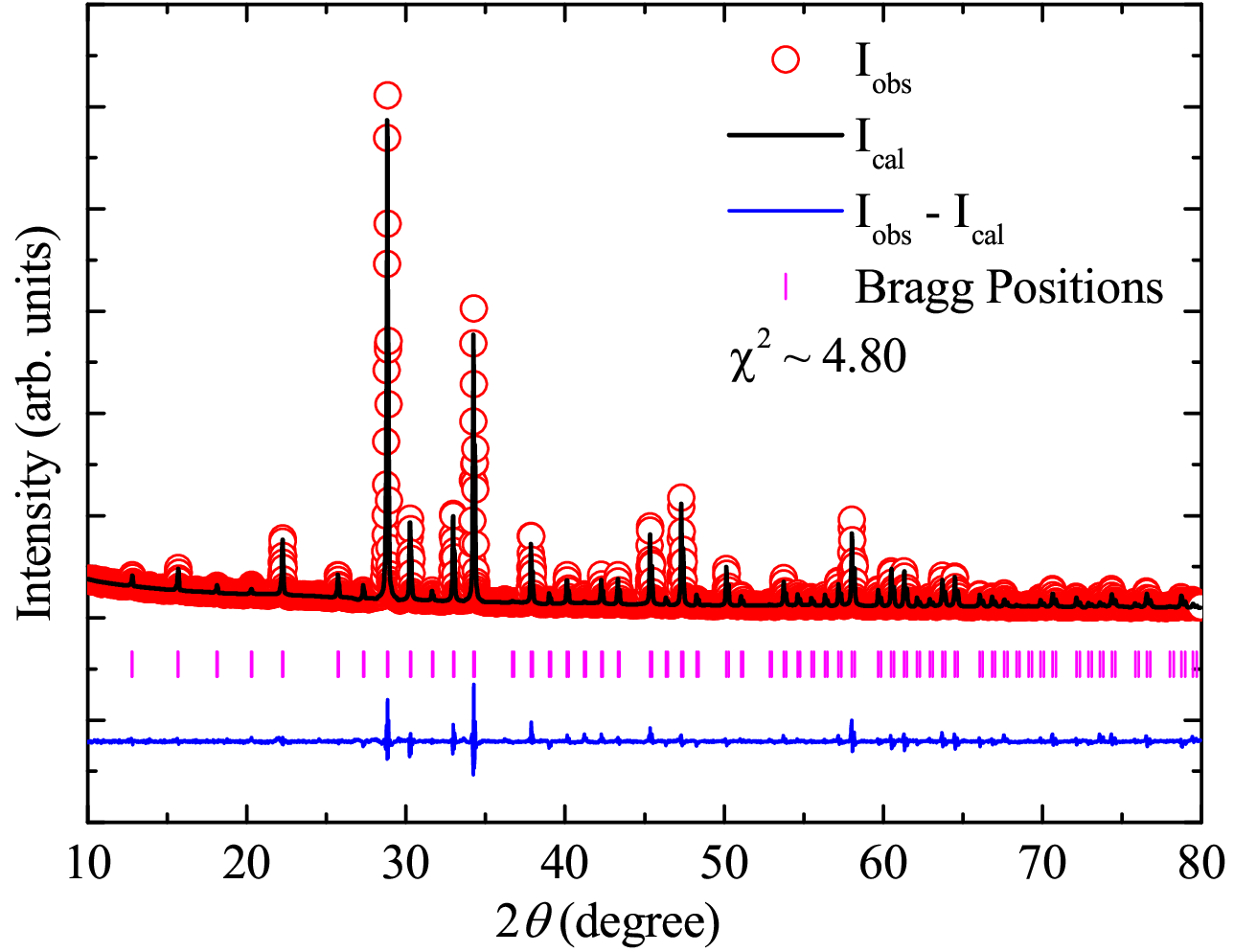}
	\caption{\label{Fig2} Room-temperature powder XRD pattern of KBaCr$_2$(PO$_4)_3$. The open circles represent the experimental data and the solid red line is the Rietveld fit. Expected Bragg-peak positions are shown as pink vertical bars and the bottom line indicates the difference in intensity between the observed and experimental data.}
\end{figure}

Polycrystalline sample of KBaCr$_2$(PO$_4$)$_3$ was prepared using the conventional solid-state method. Stoichiometric amounts of K$_2$CO$_3$ (Sigma Aldrich, 99.99\%), BaCO$_3$ (Sigma Aldrich, 99.997\%), Cr$_2$O$_3$ (Sigma Aldrich, 99.999\%), and NH$_4$H$_2$PO$_4$ (Sigma Aldrich, 99.995\%) were thoroughly grounded for several hours to make fine powder and pressed into pellets. These pellets were placed into a platinum crucible and preheated at $600~\degree$C for 12~hrs in air to release gases formed upon decomposition of the precursors. The obtained product was further grounded and pressed into pellets. These pellets were sealed in an evacuated quartz tube and fired at $1100~\degree$C for 24 hrs with several intermediate grindings. Phase purity of the sample was confirmed by powder x-ray diffraction (XRD) recorded using the PANalytical x-ray diffractometer (Cu$K_\alpha$, $\lambda_{\rm avg}= 1.5418$~\AA) at room temperature. Rietveld refinement for the powder XRD data (Fig.~\ref{Fig2}) was performed using the \texttt{FULLPROF} software package~\cite{Carvajal55} with the initial structural parameters taken from the previous report~\cite{Battle21}. The entire XRD pattern could be indexed using the cubic structure with the space group $P2_{1}3$. The resulting goodness-of-fit $\chi^2\approx 4.80$ reflects excellent sample quality. The obtained cubic lattice parameter $a\simeq 9.794$~\AA\ is in a good agreement with the literature~\cite{Battle21}.

Temperature-dependent DC magnetization ($M$) was measured using a vibrating sample magnetometer (VSM) attached to Physical Property Measurement System (PPMS, Quantum Design) in the temperature range 1.9~K to 380~K.
AC magnetization was measured in the temperature range 2~K$ \leq T \leq 15$~K by varying the frequency (100~Hz$ \leq \nu \leq 10$~KHz) in an applied AC field of $H_{\rm AC} = 5$~Oe, using the ACMS option of PPMS. Heat capacity ($C_{\rm p}$) as a function of temperature was measured on a small sintered pellet using the thermal relaxation technique in PPMS in different applied fields ranging from 0 to 9~T. Thermal conductivity measurement as a function of temperature was carried out on a small sintered cylindrical pellet using the two-probe method in PPMS, in the temperature range 2~K$ \leq T \leq 300$~K and in magnetic fields up to 9~T.

Nuclear magnetic resonance (NMR) measurements were conducted using a laboratory-built phase-coherent spin-echo pulse spectrometer over the temperatures range 1.5~K$\leq T \leq 300$~K on the $^{31}$P nucleus that has nuclear spin $I = 1/2$ and gyromagnetic ratio $\gamma_N/2\pi = 17.237$~MHz/T. The measurements were done at four different frequencies of 8.32, 17.5, 34.1, and 127.7~MHz, which correspond to the magnetic fields of about 0.5, 1, 2, and 7~T, respectively. The NMR spectra were obtained by sweeping the magnetic field at a fixed resonance frequency. For the $^{31}$P zero-shift position ($H_{\rm ref}$), a nonmagnetic reference sample H$_3$PO$_4$ was measured at room temperature. Then, the temperature-dependent NMR shift was calculated as $K(T)=[H_{\rm ref}-H(T)]/H(T)$, where $H$ is the resonance field of the sample. The nuclear spin-lattice relaxation rate ($1/T_1$) was measured at the field corresponding to the central peak position, using the standard saturation recovery technique. The nuclear spin-spin relaxation rate ($1/T_2$) was obtained by measuring the decay of the echo integral with variable spacing between the $\pi/2$ and $\pi$ pulses.

Density-functional-theory (DFT) band-structure calculations were performed in the \texttt{VASP}~\cite{vasp1,vasp2} code using the Perdew-Burke-Ernzerhof flavor of the exchange-correlation potential~\cite{pbe96}. To account for correlation effects in the Cr $3d$ shell, the mean-field DFT+$U$ method was used. The value of the on-site Coulomb repulsion $U_d$ was varied in the $2-6$\,eV range, whereas the Hund's coupling was fixed at $J_d=1$\,eV, and atomic limit was used for the double-counting correction. Experimental atomic positions from Ref.~\cite{Battle21} were utilized in all calculations. Magnetic exchange couplings of the spin Hamiltonian 
\begin{equation}
 \mathcal H=\sum_{\langle ij\rangle} J_{ij}\textbf{S}_i\textbf{S}_j,
\end{equation}
where $S=\frac32$ and the summation is over atomic pairs, were evaluated by a mapping procedure~\cite{xiang2011}.

Magnetic susceptibility for the resulting spin model was simulated using the quantum Monte-Carlo (QMC) \texttt{loop} algorithm~\cite{loop} of the \texttt{ALPS} simulation package~\cite{alps} on the finite lattice with 1728 sites and periodic boundary conditions.

\section{Results and Discussion}
\subsection{DC Magnetization}
\begin{figure*}
\includegraphics[width=\textwidth]{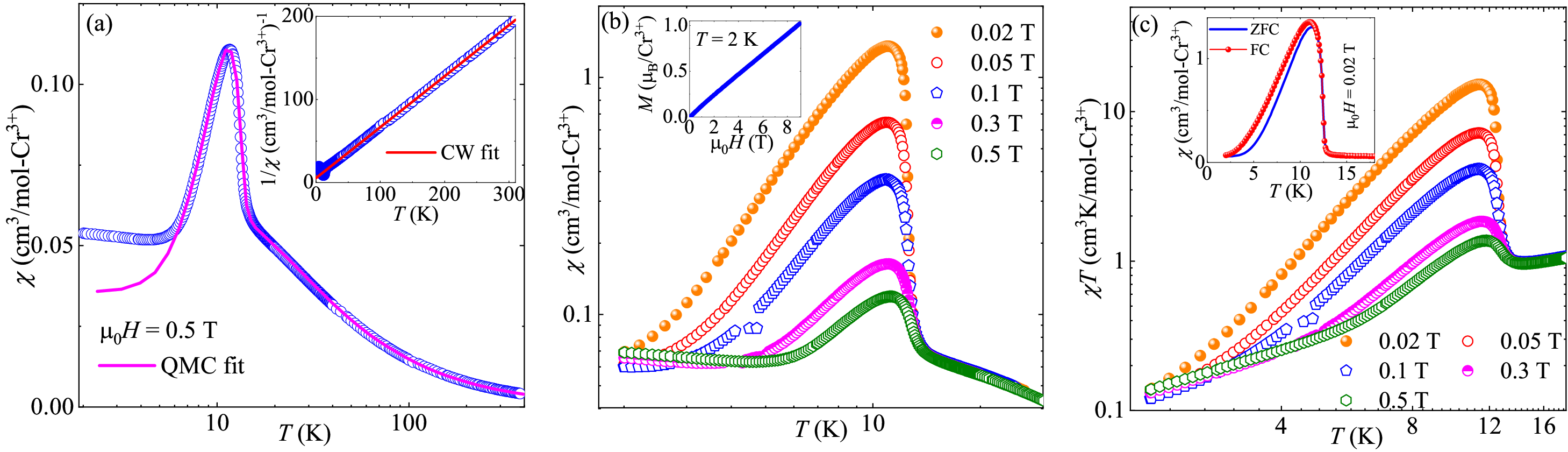}
\caption{\label{Fig3} (a) $\chi$ as a function of temperature measured in an applied magnetic field of $\mu_{0}H = 0.5$~T. The fit represents the QMC simulation for the magnetic model with the parameters from Table~\ref{tab:exchange}. Inset: Inverse susceptibility $1/\chi$ vs $T$ along with the CW fit (solid line). (b) $\chi$ vs $T$ in different magnetic fields. Inset: Magnetic isotherm ($M$ vs $H$) at $T = 2$~K. (c) $\chi T$ vs $T$ in the same fields. Inset: $\chi(T)$ measured under ZFC and FC conditions in $\mu_0H = 0.02$~T.}
\end{figure*}
Temperature variation of DC magnetic susceptibility $\chi$ ($\equiv M/H$) measured in an applied field of $\mu_{0}H = 0.5$~T is depicted in Fig.~\ref{Fig3}(a). As the temperature decreases, $\chi$ increases systematically and then features a broad peak at around $T_{\rm N1} \simeq 12$~K, indicating the onset of magnetic LRO. Magnetic parameters are determined by fitting $1/\chi$ above 160~K with the Curie-Weiss (CW) law,
\begin{equation}\label{CW}
\chi(T) = \chi_{0}+\frac{C}{(T-\theta_{\rm CW})},
\end{equation}
where $\chi_{0}$ is the $T$-independent susceptibility that includes core diamagnetism and Van-Vleck paramagnetism, $C$ is the Curie constant, and $\theta_{\rm CW}$ is the characteristic CW temperature reflecting the overall energy scale of the exchange interactions among the magnetic ions. The fit shown in the inset of Fig.~\ref{Fig3}(a) returns $\chi_{0} \simeq -1.35\times 10^{-5}$~cm$^3$/mol-Cr$^{3+}$, $C \simeq 1.63$~cm$^3$K/mol-Cr$^{3+}$, and $\theta_{\rm CW} \simeq -9.2$~K. From the value of $C$, the effective magnetic moment ($\mu_{\rm eff}$) was calculated using the relation $\mu_{\rm eff}= \sqrt{3k_{\rm B}C/N_{\rm A}}$ to be $\sim 3.61~\mu_{\rm B}$, where $N_{\rm A}$ is the Avogadro's number, $k_{\rm B}$ is the Boltzmann constant, and $\mu_{\rm B}$ is Bohr magneton. For a spin-$3/2$ system, the spin-only effective moment is $\mu_{\rm eff} = g\sqrt{S(S+1)}$ $\approx$ 3.87~$\mu_{\rm B}$, assuming the Land$\acute{e}$ $g$-factor $g = 2$. Thus, our experimentally calculated value of $\mu_{\rm eff}$ is indeed close to the expected value. The negative value of $\theta_{\rm CW}$ represents dominant antiferromagnetic (AFM) interactions. The core diamagnetic susceptibility ($\chi_{\rm core}$) caused by the orbital motion of the core electrons is calculated to be $-2.14 \times 10^{-4}$~cm$^3$/mol~\cite{Bain2008}. The Van-Vleck paramagnetic susceptibility is estimated as $\chi_{\rm VV} \simeq 2 \times 10^{-4}$~cm$^3$/mol by subtracting $\chi_{\rm core}$ from $\chi_{0}$.

The magnetic response of KBaCr$_2$(PO$_4)_3$ is rather unusual, as the initial increase in the susceptibility upon approaching $T_{\rm N1}$ is followed by its decrease toward low temperatures. This susceptibility maximum is gradually suppressed upon increasing the field [see Fig.~\ref{Fig3}(b)], as typical for systems with a ferromagnetic ordered component. On the other hand, the magnetic isotherm ($M$ vs $H$) measured at $T=2$~K [inset of Fig.~\ref{Fig3}(b)] is linear without any tendency toward saturation even at 9~T. It further does not show any remnant magnetization, thus identifying KBaCr$_2$(PO$_4)_3$ as a fully compensated antiferromagnet. The shape of $\chi T$ vs $T$ plots in Fig.~\ref{Fig3}(c) indeed imply the co-existence of FM and AFM interactions in the system~\cite{Mohanty134401}. Further, $\chi(T)$ measured under zero-field-cooled (ZFC) and field-cooled (FC) protocols in $\mu_0H = 0.02$~T [inset of Fig.~\ref{Fig3}(c)] shows only a very small difference below $T_{\rm N}$, suggesting that freezing effect play minor role in this system, in contrast to materials with structural disorder~\cite{Bag144436}.
 

\subsection{Heat Capacity}
\begin{figure*}
	\includegraphics[width=\textwidth]{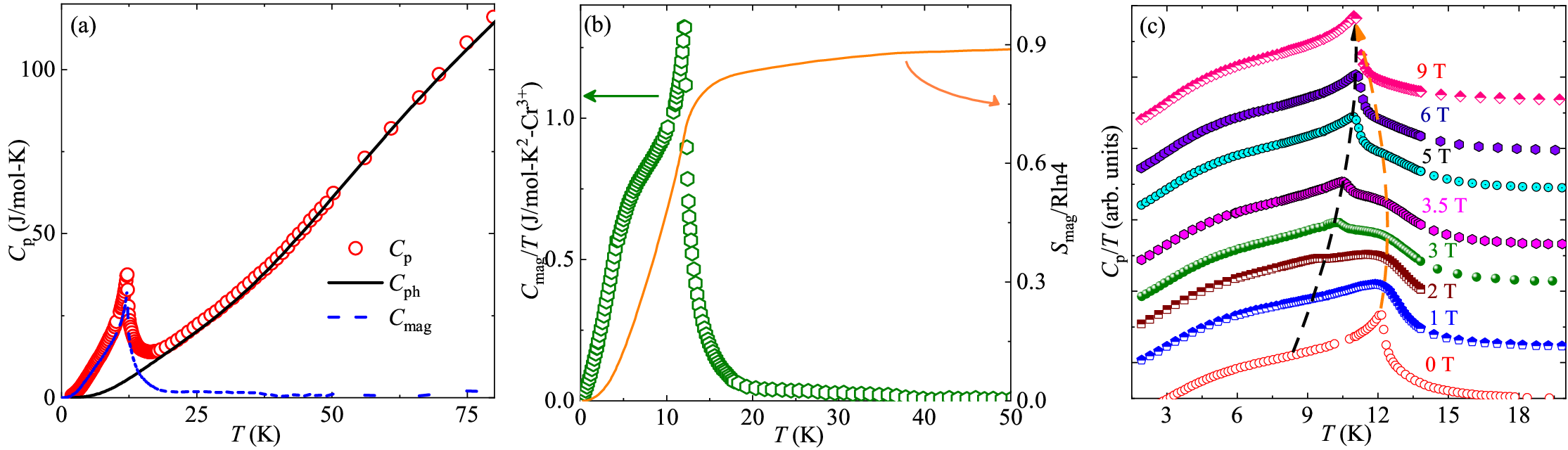}
	\caption{\label{Fig4} (a) Zero-field heat capacity ($C_{\rm p}$) vs $T$ along with the Debye-Einstein fit (solid black line). The blue dotted line represents the magnetic heat capacity ($C_{\rm mag}$). (b) $C_{\rm mag}/T$ and normalized magnetic entropy ($S_{\rm mag}/R\ln4$) vs $T$ in the left and right $y$-axes, respectively. (c) The vertically shifted $C_{\rm p}/T$ vs $T$ curves measured in different applied fields. The dashed lines trace the two transition points.}
\end{figure*}
Figure~\ref{Fig4}(a) displays the temperature-dependent heat capacity [$C_{\rm p}(T)$] measured in zero field. With lowering temperature, $C_{\rm p}$ passes through a $\lambda$-type anomaly at around $T_{\rm N1} \simeq 12.1$~K, implying the onset of magnetic LRO. Heat capacity of an insulator contains two contributions: one is phonon/lattice contribution ($C_{\rm ph}$), dominant in the high-temperature regime, while another one is magnetic contribution ($C_{\rm mag}$) that dominates over $C_{\rm ph}$ in the low-temperature region, depending on the strength of magnetic exchange couplings. In order to separate $C_{\rm mag}$ from the total heat capacity $C_{\rm p}$, we estimated $C_{\rm ph}$ by fitting the high-temperature data using a linear combination of one Debye and four Einstein terms as~\cite{Gopal2012,Mohanty134401}
\begin{equation}\label{DE}
C_{\rm ph} = f_{\rm D}C_{\rm D}(\theta_{\rm D}, T)+\sum_{i=1}^{4}g_{i}C_{\rm Ei}(\theta_{\rm Ei}, T).
\end{equation}
The first term in Eq.~\eqref{DE} is the Debye model
\begin{equation}
C_{\rm D}(\theta_{\rm D},T)=9nR\left(\frac{T}{\theta_{\rm D}}\right)^3\int_{0}^{\theta_{\rm D}/T}\frac {x^4e^x}{(e^x-1)^2} \,dx \ ,
\end{equation}
where $x=\frac{\hbar\omega}{k_{\rm B}T}$, $\omega$ is the frequency of oscillation, $R$ denotes the universal gas constant, and $\theta_{\rm D}$ is the characteristic Debye temperature. The high-energy modes of phonon vibration (optical modes) are attributed to the second term in the Eq.~\eqref{DE}, known as the Einstein term,
\begin{equation}
C_{\rm E}(\theta_{\rm E},T) = 3nR\left(\frac{\theta_{\rm E}}{T}\right)^2\frac{e^{\frac{\theta_{\rm E}}{T}}}{(e^{\frac{\theta_{\rm E}}{T}}-1)^2}.
\end{equation}
Here, $\theta_{\rm E}$ is the characteristic Einstein temperature. The coefficients $f_{\rm D}$, $g_{1}$, $g_{2}$, $g_{3}$, and $g_{4}$ are the weight factors, which take into account the number of atoms per formula units ($n$). The best fit of the zero-field $C_{\rm p} (T)$ data in the high-temperature region [black line in Fig.~\ref{Fig4}(a)] returns $\theta_{\rm D} \simeq 80$~K, $\theta_{\rm E1} \simeq 125 $~K, $\theta_{\rm E2} \simeq 248$~K, $\theta_{\rm E3} \simeq 265$~K, and $\theta_{\rm E4} \simeq 730$~K with $f_{\rm D} \simeq 0.052$, $g_{1} \simeq 0.050$, $g_{2} \simeq 0.212$, $g_{3} \simeq 0.105$, and $g_{4} \simeq 0.578$. One may notice that the sum of $f_{\rm D}$, $g_{1}$, $g_{2}$, $g_{3}$, and $g_{4}$ gives a value close to one, as expected. Finally, the high-temperature fit was extrapolated down to low-temperatures and subtracted from the experimental $C_{\rm p}(T)$ data to obtain the magnetic contribution $C_{\rm mag}(T)$.

$C_{\rm mag}(T)/T$ vs $T$ is presented in the left $y$-axis of Fig.~\ref{Fig4}(b), which shows a sharp peak at around $T_{\rm N1} \simeq 12.1$~K. Another broad hump is also observed below $T_{\rm N1}$, which is typical for systems with higher spin~\cite{Johnston094445,RNath024431}. In order to verify the estimation of magnetic contribution, the change in magnetic entropy ($\Delta S_{\rm mag}$) is evaluated by integrating $C_{\rm mag} (T)/T$ [i.e. $\Delta S_{\rm mag} = \int_{0}^{T} \frac{C_{\rm mag} (T^{\prime})}{T^{\prime}} \, dT^{\prime}$] over the whole temperature range. As shown in the right $y$-axis of Fig.~\ref{Fig4}(b), the calculated value of $\Delta S_{\rm mag}$ is $\sim 10.5$~J/mol-K which is close to 11.54~J/mol-K ($=R\ln 4$), expected for a spin-$3/2$ system.

We further measured $C_{\rm p}(T)$ in different applied magnetic fields. Increasing the field shifts $T_{\rm N1}$ towards high temperatures, but above 2\,T another peak at lower temperatures ($T_{\rm N2}$) appears. Our NMR data suggest that these two features manifest two separate transitions, even though only one clear peak is seen in the heat capacity in each field. A closer inspection of the data suggests that in fields above 2\,T, the transition at $T_{\rm N1}$ may be manifested by a shoulder. Both transitions merge at $\mu_0 H > 6$~T.
Double transitions have been reported in many frustrated magnets and can be attributed to the magnetic anisotropy~\cite{Ranjith014415,Mohanty104424,Lal014429}.

\subsection{AC Magnetization}
\begin{figure}
	\includegraphics[width=\columnwidth]{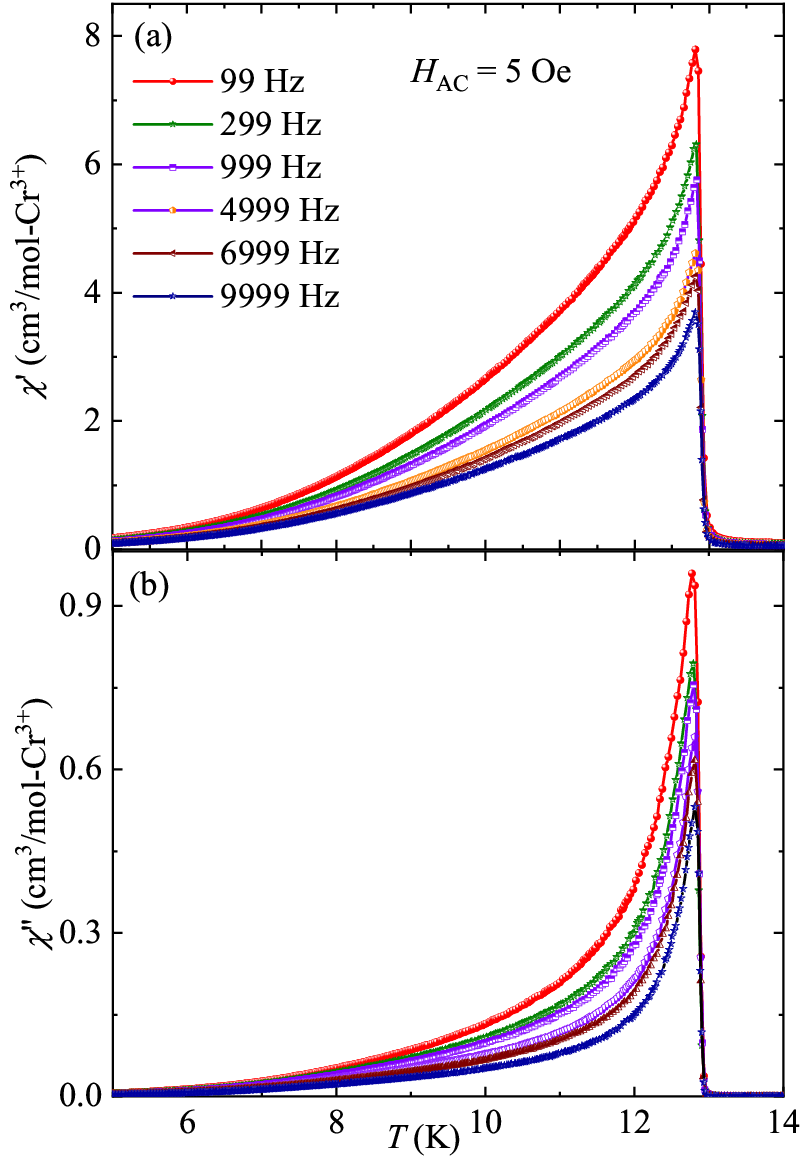}
	\caption{\label{Fig5} (a) Real part of the AC susceptibility ($\chi^{\prime}$) vs $T$ and (b) Imaginary part of the AC susceptibility ($\chi^{\prime\prime}$) vs $T$, measured at different frequencies.}
\end{figure}
To explore the possibility of the double magnetic transition in zero field, we measured magnetic susceptibility in the applied AC field of $H_{\rm ac} = 5$~Oe at different frequencies. Figure~\ref{Fig5}(a) and (b) present the temperature variation of the real part ($\chi^{\prime}$) and the imaginary part ($\chi^{\prime\prime}$), respectively. Both $\chi^{\prime}$ and $\chi^{\prime\prime}$ show a sharp peak at $T_{\rm N1} \simeq 12.8$~K, which is close to the anomaly observed in the zero-field $C_{\rm p}(T)$ data. The peak position is independent of frequency, suggesting the absence of freezing effects. No feature associated with the second transition at $T_{\rm N2}$ has been observed. 

\subsection{Thermal Conductivity}
\begin{figure}
	\includegraphics[width=\columnwidth]{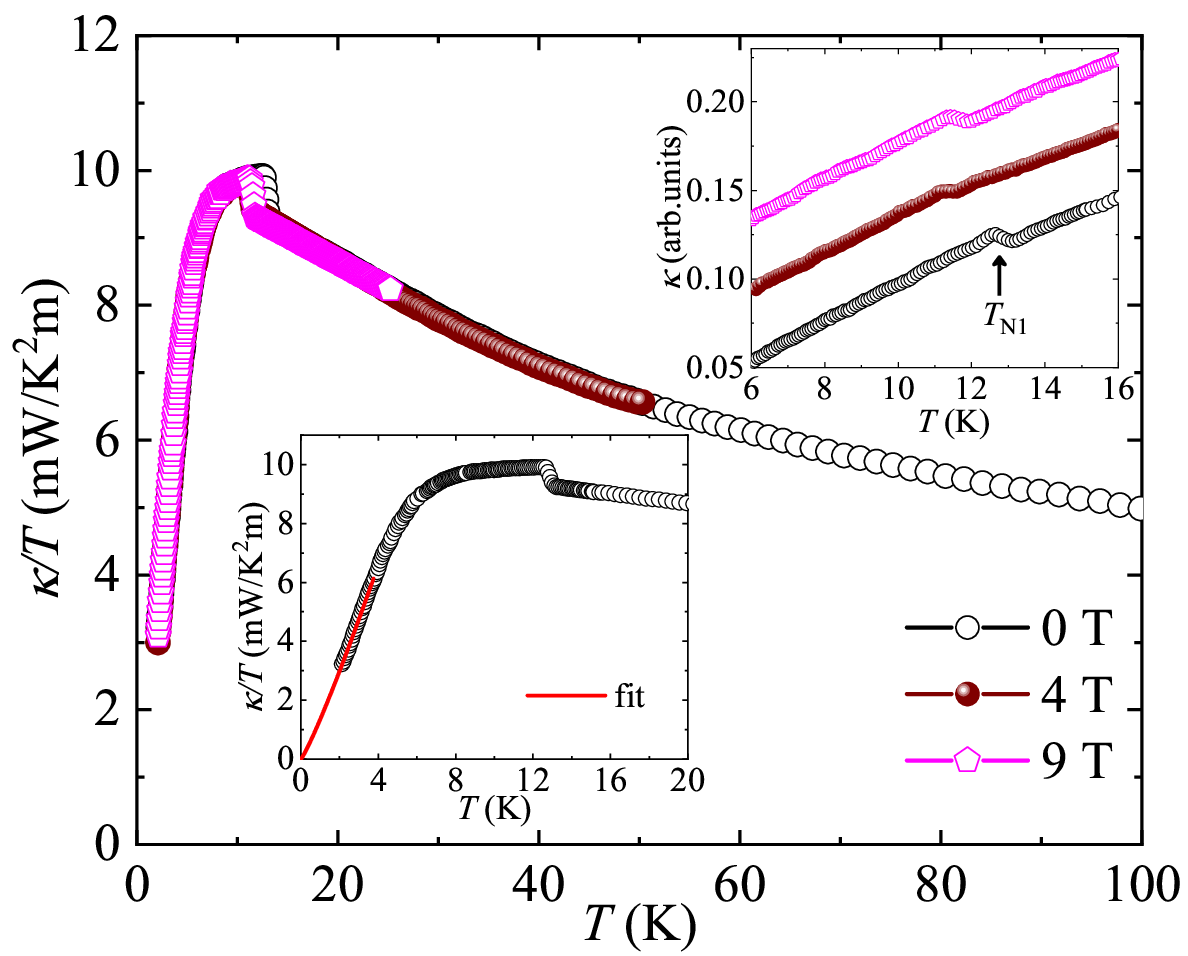}
	\caption{\label{Fig6} $\kappa/T$ vs $T$ measured in three different applied fields. Upper inset: Vertically translated $\kappa$ vs $T$ data around $T_{\rm N1}$ to highlight the transition. Lower inset: Zero-field $\kappa/T$ vs $T$ in the low-temperature regime. The solid line is the fit as described in the text.}
\end{figure}
In contrast to the heat-capacity data, thermal conductivity is insensitive to nuclear degrees of freedom~\cite{Yamashita1246} and serves as a useful probe of itinerant spin excitations at low temperatures.
Figure~\ref{Fig6} shows the temperature-dependent thermal conductivity divided by temperature ($\kappa/T$) measured in several applied fields. As temperature decreases, $\kappa(T)$ increases slowly and exhibits a step-like feature at around $T_{\rm N1} \simeq 12.5$~K, related to the magnetic transition. No anomaly associated with the second transition at $T_{\rm N2}$ could be detected in zero field. As the magnetic field is applied, the step shifts with temperature [upper inset of Fig.~\ref{Fig6}], similar to that observed in the $C_{\rm p}(T)$ data.

Zero-field thermal conductivity at low temperatures ($T < T_{\rm N1}$) was fitted by $\kappa/T = a + bT^{\alpha - 1}$~\cite{Yamashita1246,Huang2022}. Here, the first term represents the possible contribution of itinerant fermionic magnetic excitations, while the second term stands for conventional magnons and lattice vibrations. As shown in the lower inset of Fig.~\ref{Fig6}, the fit below 3.5~K yields $a=0$ with $\alpha \simeq 2.14$. 
One generally expects $\alpha=3$ for both phonons and magnons in 3D. The lower value of the exponent may be due to phonon reflections at the sample surfaces or due to magnon-phonon scattering~\cite{Guang184423}. 

\subsection{$^{31}$P NMR}
\begin{figure}
	\includegraphics[width=\columnwidth]{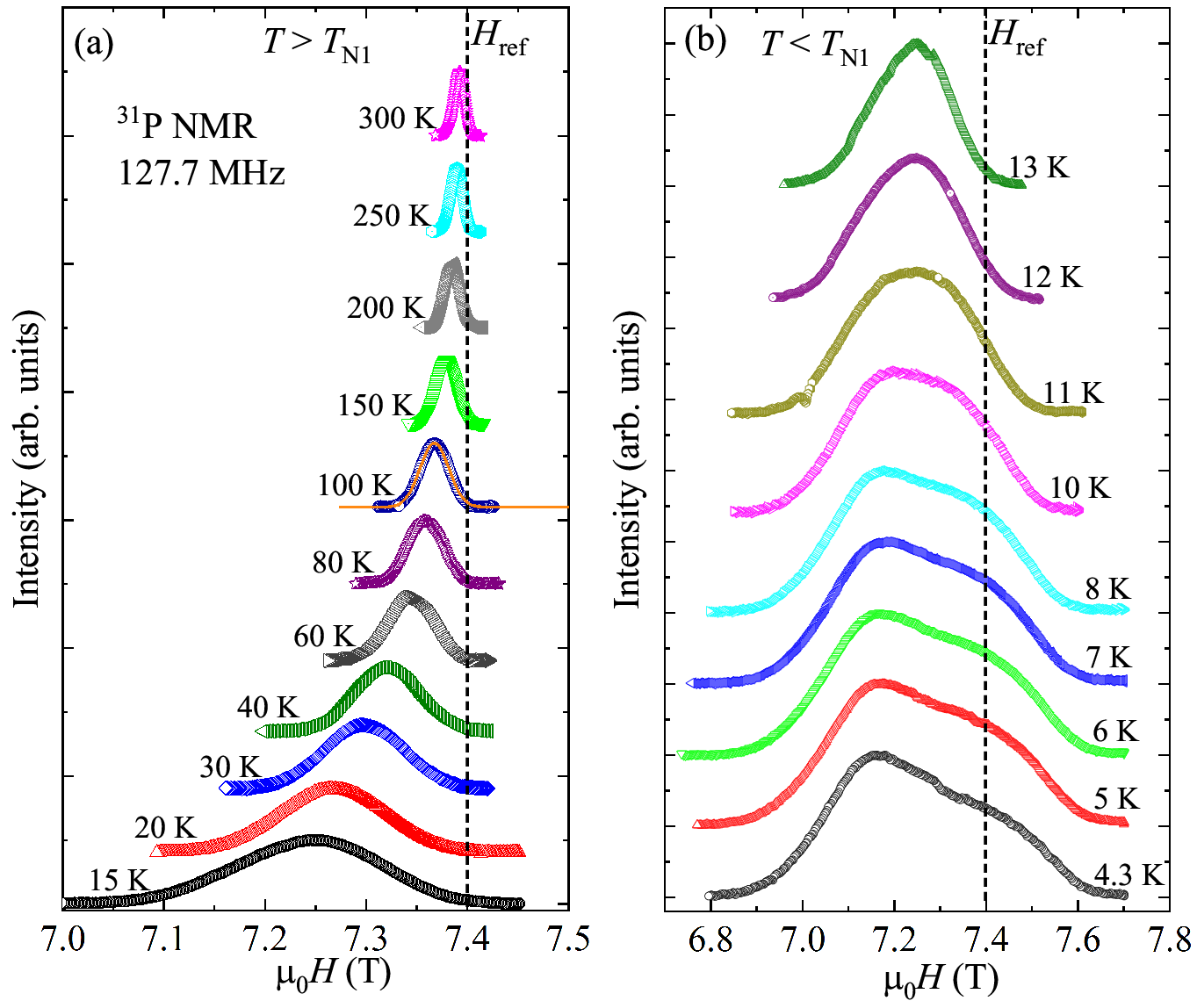}
	\caption{\label{Fig7} Temperature-dependent $^{31}$P NMR spectra, measured in an applied frequency of $127.7$~MHz (a) above $T_{\rm N1}$ and (b) below $T_{\rm N1}$. The vertical dashed line represents the nonmagnetic $^{31}$P reference field.}
\end{figure}

\begin{figure}
	\includegraphics[width=\columnwidth]{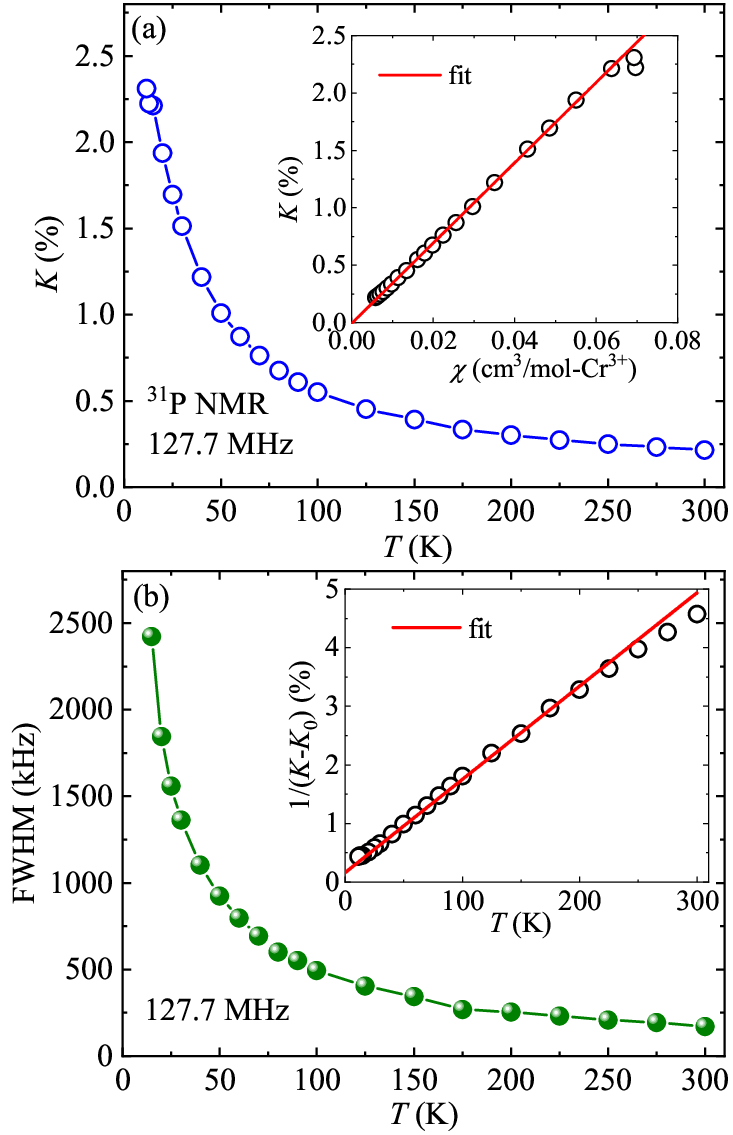}
	\caption{\label{Fig8} (a) Temperature-dependent $^{31}$P NMR shift measured at 127.7~MHz. Inset: $K$ vs $\chi$ (measured at 2~T). The solid line is the linear fit. (b) Temperature-dependent full-width at half-maximum (FWHM). Inset: $1/(K-K_0)$ vs $T$ and the solid line is the fit as described in the text.}
\end{figure}
We further use NMR to investigate both static and dynamic properties of the material. The KBaCr$_2$(PO$_4)_3$ structure features only one $^{31}$P site. The Cr(1)O$_6$ and Cr(2)O$_6$ octahedra are connected via the PO$_4$ tetrahedra [see Fig.~\ref{Fig1}(a)]. Therefore, the $^{31}$P site serves as a convenient local probe of the Cr$^{3+}$ magnetism.

\subsubsection{$^{31}$P NMR Spectra}
Field-sweep $^{31}$P NMR spectra measured at the radio frequency of 127.7~MHz are shown in Fig.~\ref{Fig7}. The single spectral line is consistent with the unique $^{31}$P site. At high temperatures, the spectrum is narrow. As the temperature decreases, the line broadens and becomes asymmetric, as typical of polycrystalline samples due to the effects of asymmetric hyperfine coupling and/or anisotropic susceptibility~\cite{Yogi024413}. The line position also shifts towards lower magnetic fields with decreasing temperature. To visualize temperature-induced changes in the line broadening and line shift, each spectrum is normalized using the peak amplitude and vertically offset in Fig.~\ref{Fig7}.

The NMR shift $K(T)$ plotted in Fig.~\ref{Fig8}(a) strongly resembles the bulk $\chi(T)$ data, thus proving the absence of any significant impurity contribution because the NMR shift probes local magnetic susceptibility ($\chi_{\rm spin}$).
Their relation can be expressed as
\begin{equation}
    K (T) = K_0 + \frac{A_{\rm hf}}{N_{\rm A}}\chi_{\rm spin}(T)
\end{equation}
where $K_0$ is the temperature-independent chemical shift, and $A_{\rm hf}$ is the hyperfine coupling constant between the $^{31}$P nuclear spin and Cr$^{3+}$ electronic spins. 
The linear fit of $K$ vs $\chi$ in the inset of Fig.~\ref{Fig8}(a) returns $K_0 \simeq -0.009$~\% and $A_{\rm hf} \simeq 1970.6$~Oe/$\mu_{\rm B}$. Temperature variation of $1/(K-K_0)$ is shown in the inset of Fig.~\ref{Fig8}(b). The data in the $T$-range from 50~K to 250~K are fitted using the CW law, $K = K_{0} + B/(T-\theta_{\rm CW}^{K})$. The effective moment determined from the $B$ value is $\mu_{\rm eff}^{K} \simeq 3.77$~$\mu_{\rm B}$ in a very good agreement with the expected value for spin-$3/2$~\cite{Nath024418}. Similarly, the resulting value of $\theta_{\rm CW}^{K} \simeq -10$~K is also in close agreement with the results of the $\chi(T)$ analysis. The full-width at half-maximum (FWHM) [Fig.~\ref{Fig8}(b)] increases slowly on cooling and diverges on approaching $T_{\rm N1}$, indicating the growth of internal static fields. 

\subsubsection{$^{31}$P spin-lattice relaxation rate $1/T_1$}
\begin{figure}
	\includegraphics[width=\columnwidth]{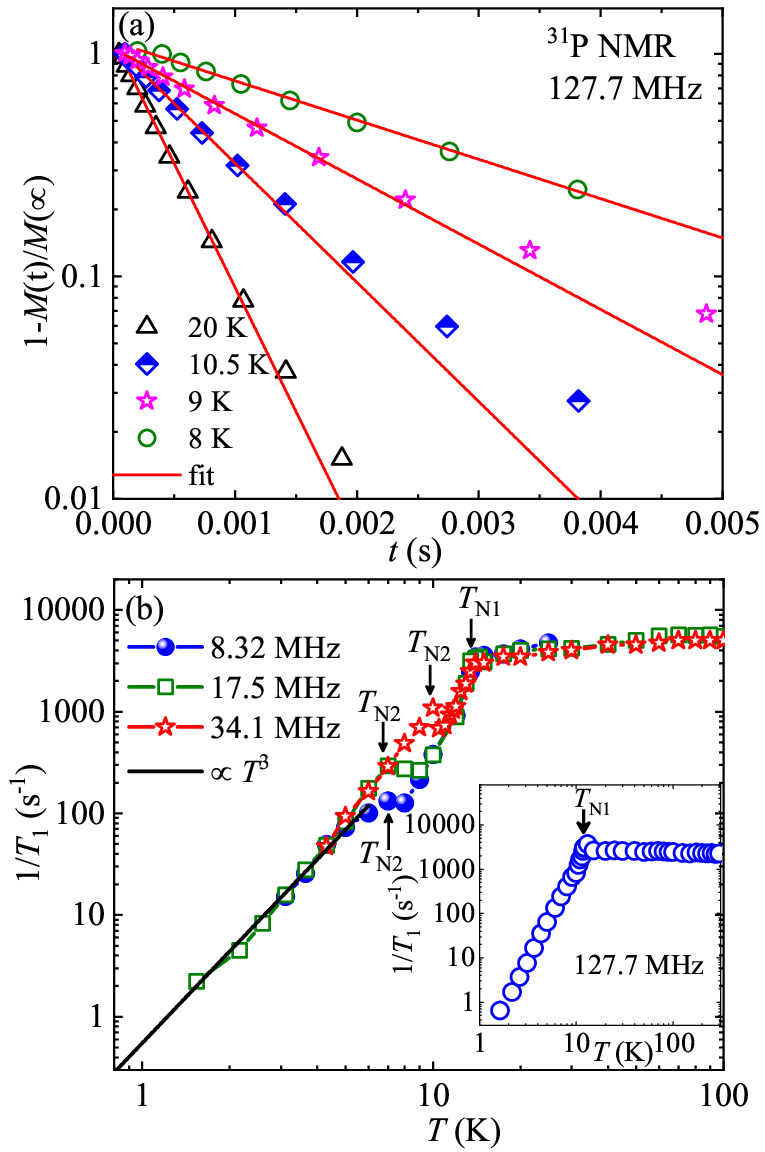}
	\caption{\label{Fig9} (a) Longitudinal magnetization recovery curves measured at 127.7~MHz for four selected temperatures. Solid lines are the fits using Eq.~\eqref{recovery}. (b) Temperature variation of the $^{31}$P spin-lattice relaxation rate, $1/T_{1}$, at three different frequencies, 8.32~MHz, 34.1~MHz, and 17.5~MHz. The arrows mark two successive magnetic transitions at $T_{\rm N1}$ and $T_{\rm N2}$. The solid line represents the $T^{3}$ fit below $T_{\rm N2}$. Inset: $1/T_{1}$ vs $T$ at 127.7~MHz.}
\end{figure}
The $^{31}$P spin-lattice relaxation rate $1/T_1$ was measured at the field corresponding to the central peak position down to 1.5~K at four different frequencies 8.32~MHz, 17.5~MHz, 34.1~MHz, and 127.7~MHz. Typically, for an $I = 1/2$ nucleus, the recovery of the longitudinal magnetization should follow a single exponential function
\begin{equation}
\label{recovery}
    1-\frac{M(t)}{M(\infty)} = Ae^{-t/T_1},
\end{equation}
where $M(t)$ is the nuclear magnetization at a time $t$ after the saturation pulse, and $M(\infty)$ is the equilibrium value of magnetizaton. The recovery curves for 127.7~MHz at four different temperatures along with the fits using the Eq.~\eqref{recovery} are shown in Fig.~\ref{Fig9}(a).

The extracted temperature variation of $1/T_{1}$ is depicted in Fig.~\ref{Fig9}(b) for 8.32, 17.5, and 34.1~MHz. At high temperatures ($T \geq 20$~K), $1/T_{1}$ is almost temperature-independent as the moments are fluctuating fast and randomly in the paramagnetic regime~\cite{Moriya23}. At low temperatures, two anomalies in $1/T_1(T)$ appear: a sharp kink at $T_{\rm N1}$ is followed by a weak anomaly $T_{\rm N2}$ that confirms two successive magnetic transitions in finite magnetic fields. The anomaly at $T_{\rm N1} \simeq 13.5$~K remains almost frequency-independent, whereas the one at $T_{\rm N2}$ moves toward higher temperatures with increasing frequency ($T_{\rm N2} \sim 7$~K, $\sim 7.5$~K, and $\sim 10$~K for 8.32~MHz, 17.5~MHz, and 34.1~MHz, respectively). These results are consistent with those found from the heat capacity data. 
Below $T_{\rm N2}$, $1/T_{1}$ decreases toward zero as a result of the melting of the critical fluctuations in the ordered state and scattering of magnons by nuclear spins~\cite{Nath214430,Belesi184408}. The $1/T_{1}$ vs $T$ measured at 127.7~MHz [inset of Fig.~\ref{Fig8}(b)] exhibits only one peak at around $T_{\rm N1} \simeq 13.5$~K suggesting that the two transitions are merged into one in higher fields.

In the ordered state ($T < T_{\rm N}$) and for $T \gg \Delta/k_{\rm B}$ ($\Delta$ is the energy gap in the acoustic magnon spectrum), $1/T_{1}$ either obeys the $T^3$ behavior due to a two-magnon Raman process or the $T^5$ behavior due to a three magnon process~\cite{Beeman359}. Clearly, the $T^3$ behavior of $1/T_1$ below $T_{\rm N2}$ [see Fig.~\ref{Fig9}(b)] suggests that the relaxation is dominated by two-magnon processes in this compound. Similar behavior is also reported in several other low-dimensional or frustrated magnets~\cite{Nath214430,Mohanty104424}. 

From the high-temperature part of $1/T_{1}$, one can estimate the effective exchange coupling between the Cr$^{3+}$ moments as~\cite{Moriya23,Guchhait024426},
\begin{equation}
    {\left(\frac{1}{T_1}\right)}_{T \rightarrow\infty} = \frac{(\gamma_{\rm N}g\mu_{\rm B})^2\sqrt{2\pi}z^{\prime}S(S+1)}{3 \omega_{\rm ex}} {\left(\frac{A_z}{z^\prime}\right)}^2.
\end{equation}
 Here, $\omega_{\rm ex} = (J_{\rm max}k_{\rm B}/\hbar)\sqrt{2zS(S+1)/3}$ is the Heisenberg exchange energy, $z=6$ is the number of nearest-neighbor spins of each Cr$^{3+}$ ion assuming the trillium-lattice geometry, and $z^{\prime}=4$ represents the number of nearest-neighbor Cr$^{3+}$ spins for a given P site. 
Using the relevant parameters, $A_{z} \simeq 1970.6$~Oe/$\mu_{\rm B}$, $\gamma_{\rm N} = 108.28$~rad s$^{-1}$ Oe$^{-1}$, $g = 2$, $S = 1/2$, and the high-temperature (150~K) relaxation rate of ${\left(\frac{1}{T_1}\right)}_{T\rightarrow\infty} \simeq 2325$~s$^{-1}$, the magnitude of the maximum exchange coupling constant is calculated to be $J_{\rm max}/k_{\rm B} \simeq 1.21$~K. This is consistent with the average value of the exchange couplings estimated from the band-structure calculations.

\subsubsection{$^{31}$P spin-spin relaxation rate $1/T_2$}
\begin{figure}
	\includegraphics[width=\columnwidth]{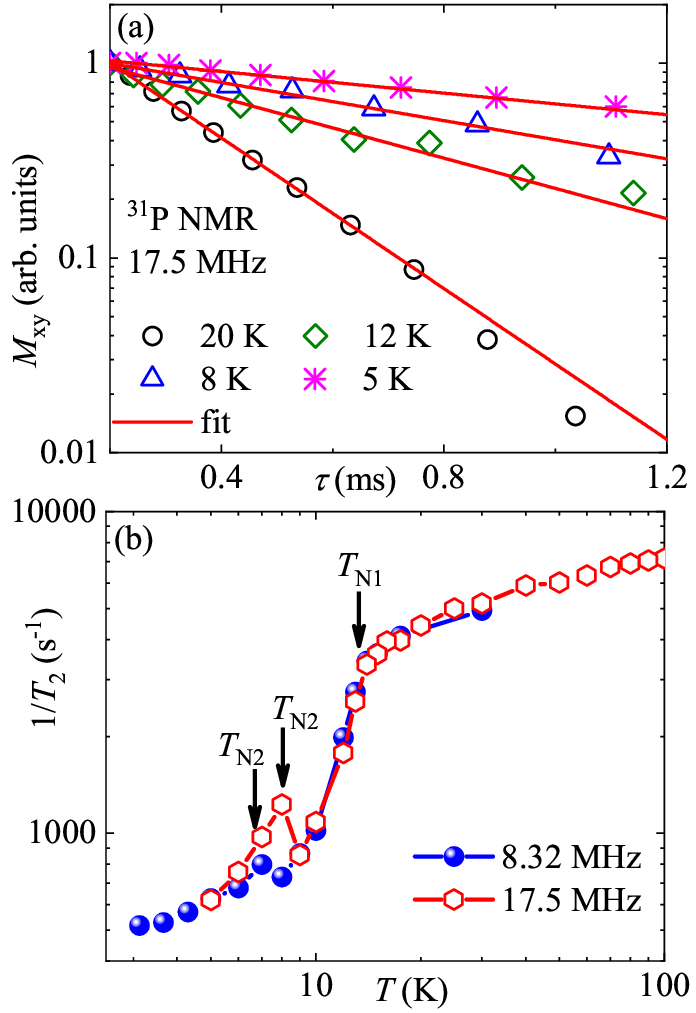}
	\caption{\label{Fig10} (a) Transverse magnetization recovery curves as a function of $\tau$ at four different temperatures measured at 17.5~MHz. (b) $1/T_2$ vs $T$. The downward arrows mark two magnetic transitions.}
\end{figure}
 The $^{31}$P spin-spin relaxation rate ($1/T_2$) was recorded by monitoring the decay of the transverse magnetization ($M_{\rm xy}$) after a $\pi/2-\tau-\pi$ pulse sequence as a function of the pulse separation time $\tau$. The $M_{\rm xy}$ can be fitted using the following exponential function
 \begin{equation}\label{SSR}
     M_{\rm xy} = M_0e^{-2\tau/T_2}.
 \end{equation}
The recovery curves at 17.5~MHz for a few selected temperatures are presented in Fig.~\ref{Fig10}(a). The obtained $1/T_2$ from the fit using Eq.~\eqref{SSR} as a function of temperature is shown in Fig.~\ref{Fig10}(b) for two different frequencies. At low temperatures, it exhibits two anomalies for both frequencies. The anomaly at $T_{\rm N1} \simeq 13.5$~K remains unchanged, whereas $T_{\rm N2}$ shifts toward higher temperatures ($T_{\rm N2} \simeq 7$~K and 7.5~K for 8.32~MHz and 17.5~MHz, respectively) with increasing frequency, similar to that observed in the $1/T_1$ measurements.

\subsubsection{Low-$T$ spectra}
\begin{figure}
	\includegraphics[width=\columnwidth]{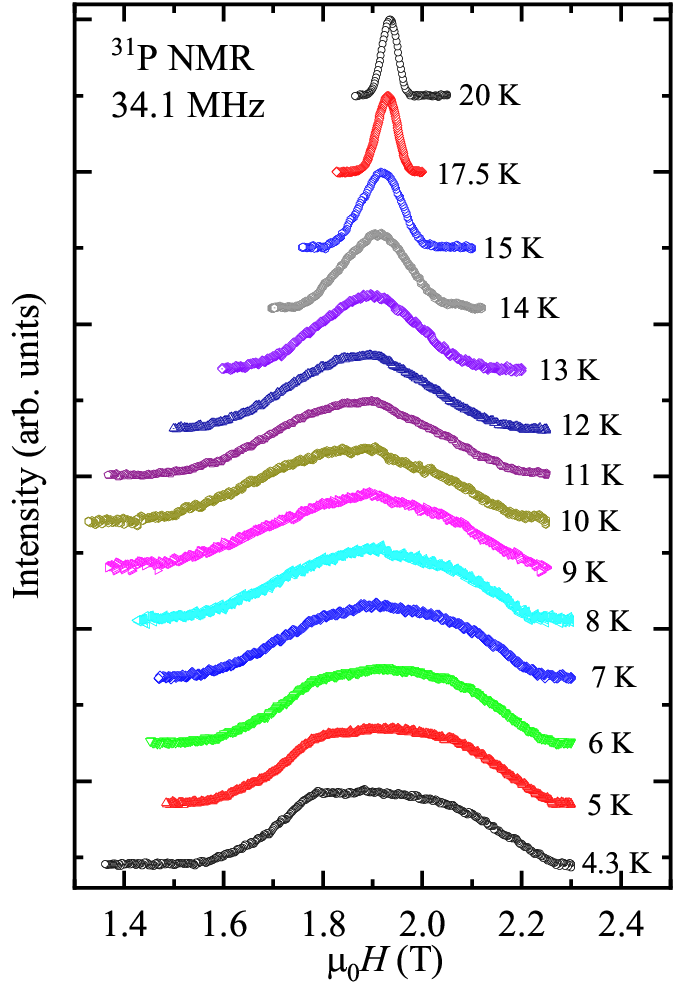}
	\caption{\label{Fig11} $^{31}$P NMR spectra in the low-temperature regime measured at 34.1~MHz.}
\end{figure}
In order to understand the nature of transitions, we measured $^{31}$P NMR spectra in the low-temperature regime and in a lower frequency of 34.1~MHz. Usually NMR spectrum becomes broader with increasing magnetic field in a power sample if the distribution of the hyperfine fields originating from the anisotropy in the hyperfine coupling constant and/or the  magnetic susceptibility.
Therefore, measurements in a lower field (and, correspondingly, at a lower frequency) may provide the intrinsic NMR line shape~\cite{Drain1380,Nath174436}. As shown in Fig.~\ref{Fig11}, the spectral width at 34.1~MHz is reduced drastically as compared to 127.7~MHz. Further, the NMR line is found to broaden systematically with decreasing temperature. Below $T_{\rm N1}$, it broadens abruptly due to the development of static internal magnetic field in the LRO state~\cite{Ranjith014415}. At very low temperatures, though, no significant line broadening is observed but the line width remains almost unchanged below about 10~K, which matches with the value of $T_{\rm N2}$. From the shape of the powder NMR spectra, it is often possible to deduce the nature of the ordered phase. For instance, a nearly rectangular line shape is anticipated for a commensurate magnetic ordering, while an incommensurate spin-density-wave state would produce a triangular powder pattern~\cite{Ranjith014415,Kikuchi2660,Yamada1751,Kontani672}. Interestingly, the shape of the powder pattern in Fig.~\ref{Fig11} below $T_{\rm N1}$ is neither rectangular nor completely triangular in shape. By contrast, collinear LRO has been revealed by neutron diffraction in zero field~\cite{Battle21}. One possible explanation is that some departure from collinearity happens below $T_{\rm N2}$ when magnetic field is applied.

\begin{figure}
	\includegraphics[width=\columnwidth]{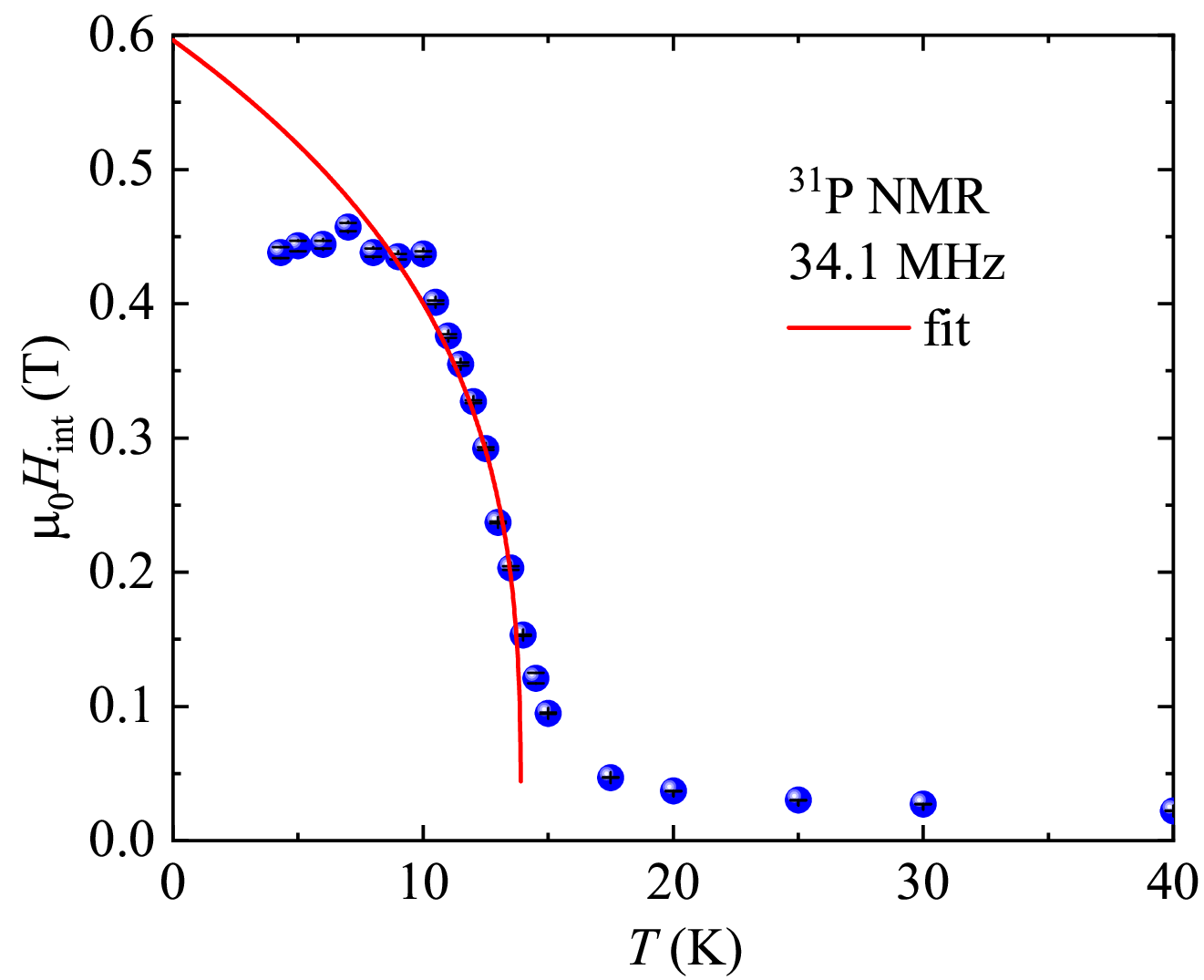}
	\caption{\label{Fig12} Temperature-dependent internal magnetic field $H_{\rm int}$ obtained from the $^{31}$P NMR spectra measured at 34.1~MHz in the ordered state. The solid line is the fit using Eq.~\eqref{Hint}.}
\end{figure}
The internal magnetic field $H_{\rm int}$, which is proportional to the Cr$^{3+}$ sublattice magnetization, was determined by taking FWHM of the spectra measured at 34.1~MHz. Below $T_{\rm N1}$, this internal field increases much faster than the mean-field prediction (Fig.~\ref{Fig12}) and abruptly saturates below $T_{\rm N2}$. The critical exponent of the order parameter ($\beta$) is estimated from
\begin{equation}\label{Hint}
    H_{\rm int}(T) = H_0{\left(1-\frac{T}{T_{\rm N1}}\right)}^\beta.
\end{equation}
The fit returns $\beta \simeq 0.31$, $\mu_0 H_0 \simeq 0.59$~T, and $T_{\rm N1} \simeq 13.9$~K. The $\beta$ value is consistent with any of the 3D universality classes (Heisenberg, Ising, or XY) and reflects 3D nature of the ordering at $T_{\rm N1}$~\cite{Nath214430}.

\subsection{Microscopic magnetic model}
\begin{figure*}
	\includegraphics[width=\textwidth]{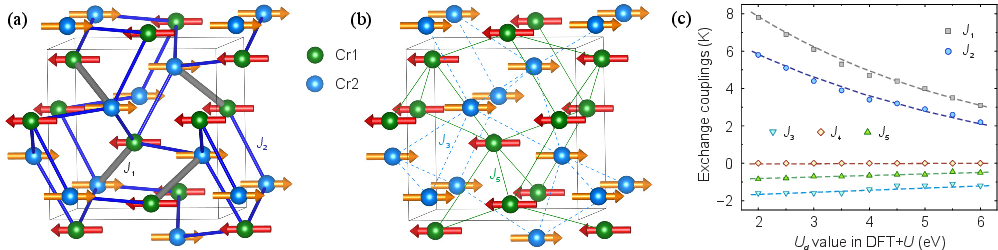}
	\caption{\label{Fig13} (a,b) Spin lattice of KBaCr$_2$(PO$_4)_3$ and the $\mathbf k=0$ magnetic order~\cite{Battle21} that satisfies both AFM couplings $J_1$, $J_2$ as well as FM coupling $J_3$, $J_5$. The spin direction $[100]$ is chosen arbitrarily. (c) Exchange couplings calculated for different $U_d$ values in DFT+$U$.}
\end{figure*}
The calculated exchange couplings for KBaCr$_2$(PO$_4)_3$ are listed in Table~\ref{tab:exchange}. To simplify the comparison to other langbeinite compounds, we use the same notation of $J_1-J_5$ as in the case K$_2$Ni$_2$(SO$_4)_3$~\cite{Ivica157204} and, therefore, do not follow the hierarchy of the Cr--Cr distances in KBaCr$_2$(PO$_4)_3$. Interestingly, the coupling regimes of the Ni and Cr compounds are entirely different from each other. K$_2$Ni$_2$(SO$_4)_3$ features large AFM $J_4$ and $J_5$. By contrast, the dominant terms in KBaCr$_2$(PO$_4)_3$ are $J_1$ and $J_2$, both AFM. On the other hand, $J_4$ is negligible, whereas $J_3$ and $J_5$ are weakly FM. This coupling regime eliminates the frustration of the double-trillium spin lattice. A simple two-sublattice $\mathbf k=0$ order with opposite spin directions on Cr1 and Cr2 satisfies all exchange couplings in this material [Fig.~\ref{Fig12}(a,b)]. Such a ground state has been indeed observed experimentally by neutron diffraction~\cite{Battle21}. We further checked the dependence of magnetic couplings on the $U_d$ parameter of DFT+$U$ and found that this non-frustrated coupling regime persists within the realistic windows of the $U_d$ values [Fig.~\ref{Fig12}(c)]. 

The best agreement with the experimental Curie-Weiss temperature is found at $U_d=5$\,eV ($\theta_{\rm CW}=-9.2$\,K averaged over the two Cr sites), which is slightly above the typical $U_d$ values of $3-4$\,eV employed in the previous studies of Cr$^{3+}$ magnets~\cite{janson2013,janson2014}. We further simulated magnetic susceptibility for the spin lattice with the exchange couplings $J_1-J_5$ using QMC. The values listed in Table~\ref{tab:exchange} allow a very accurate description of the experimental susceptibility data and show a good match with the DFT results.

\begin{table}
\caption{\label{tab:exchange}
The Cr--Cr distances $d_i$ (in\,\r A) and exchange couplings $J_i$ (in\,K) for KBaCr$_2$(PO$_4)_3$. The $J_i^{\rm DFT}$ values are obtained from the DFT+$U$ calculations with $U_d=5$\,eV and $J_d=1$\,eV, whereas the $J_i^{\,\rm QMC}$ values are the best fit to the experimental magnetic susceptibility. The notation of Cr1 and Cr2 follows Ref.~\cite{Battle21}.}
\begin{ruledtabular}
\begin{tabular}{ccc@{\hspace{1cm}}rr}
       &          &  $d_i$ & $J_i^{\rm DFT}$ & $J_i^{\,\rm QMC}$ \\
 $J_1$ & Cr1--Cr2 & 4.496  &    4.0          &     3.4     \\
 $J_2$ & Cr1--Cr2 & 4.919  &    2.9          &     2.5     \\
 $J_3$ & Cr2--Cr2 & 6.026  &  $-1.2$         &    $-0.9$   \\
 $J_4$ & Cr1--Cr2 & 5.971  &    0.0          &     0.0     \\
 $J_5$ & Cr1--Cr1 & 6.084  &  $-0.6$         &    $-0.4$   \\
\end{tabular}
\end{ruledtabular}
\end{table}

Exchange fields on the Cr1 and Cr2 sites are different by virtue of the difference between $J_3$ and $J_5$: compare $H_1\sim J_1+3J_2+3J_4+6J_5=10.9$\,K and $H_2\sim J_1+3J_2+3J_4+6J_3=5.5$\,K (at $U_d=5$\,eV). The two sublattices will thus show different temperature dependence of the magnetization, resulting in an incomplete compensation in the vicinity of $T_N$. This explains the abrupt increase in the magnetic susceptibility right below $T_N$ with the eventual formation of the fully compensated state at $T\ll T_N$, where both sublattices develop equal magnetizations. Such a behavior is well reproduced by our QMC simulations [Fig.~\ref{Fig3}(a)]. Similar physics has been reported, for example, in Mn$_2$Mo$_3$O$_8$ where two magnetic sublattices are also formed by two distinct structural sublattices of the material~\cite{lippold1990,kurumaji2017}.

A closer inspection of the crystal structure reveals that the main exchange pathways in KBaCr$_2$(PO$_4)_3$, $J_1$ and $J_2$, are those mediated by double bridges of the PO$_4$ tetrahedra. This coupling regime is typical for V$^{4+}$ phosphates where double-tetrahedral bridges result in stronger superexchange couplings compared to the single-tetrahedron bridges~\cite{tsirlin2011}. A similar coupling regime is indeed expected in Cr$^{3+}$ phosphates because magnetic orbitals are also $t_{2g}$ in nature. By contrast, K$_2$Ni$_2$(SO$_4)_3$ exhibits a very different microscopic scenario because its magnetic orbitals belong to the $e_g$ manifold. These orbitals feature a $\sigma$-overlap with oxygen $p$-orbitals and boost the couplings $J_4$ and $J_5$ that run via the single PO$_4$ tetrahedron each. We also note that only $J_1$ and $J_2$ show the typical $1/U_d$ dependence [Fig.~\ref{Fig12}(c)] expected for magnetic interactions dominated by superexchange. On the other hand, $J_3$ and $J_5$ remain almost unchanged when $U_d$ is increased, so their ferromagnetic nature is due to the weak potential exchange in the absence of any significant contribution from superexchange.

\section{Summary}
In summary, we have shown that KBaCr$_2$(PO$_4$)$_3$ manifests the nonfrustrated coupling regime caused by the interplay of ferromagnetic interactions within each Cr sublattice and antiferromagnetic interactions between these sublattices. Different exchange fields on the two sublattices give rise to an incomplete compensation and the prominent susceptibility maximum right below $T_{\rm N1}$, whereas at $T\ll T_{\rm N1}$ the system tends toward a fully compensated antiferromagnet. Despite the absence of frustration, KBaCr$_2$(PO$_4$)$_3$ reveals further interesting features, most notably, the second magnetic transition at $T_{\rm N2}$ manifested by a peak in the $^{31}$P spin-lattice relaxation and spin-spin relaxation rates. This second transition is clearly observed in applied fields only and may correspond to a small deviation of the magnetic order from the collinear order observed in zero field. 

In the broader context of langbeinite-type compounds, KBaCr$_2$(PO$_4$)$_3$ gives an instructive example of the frustration release on the trillium lattice. All of the magnetic couplings in the langbeinite structure are long-range in nature, but they are not necessarily antiferromagnetic and, therefore, not all members of this structural family are magnetically frustrated. We also showed that the nature of the $3d$ ion determines leading magnetic interactions in the langbeinite structure.


\acknowledgments
RK and RN would like to acknowledge SERB, India, for financial support bearing sanction Grant No.~CRG/2019/000960. Work at the Ames National Laboratory was supported by the U.S. Department of Energy, Office of Science, Basic Energy Sciences, Materials Sciences and Engineering Division. The Ames National Laboratory is operated for the U.S. Department of Energy by Iowa State University under Contract No.~DEAC02-07CH11358. The authors gratefully acknowledge the computing time made available to them on the high-performance computer at the NHR Center of TU Dresden. This center is jointly supported by the Federal Ministry of Education and Research and the state governments participating in the NHR (www.nhr-verein.de/unsere-partner).


%

\end{document}